# Recursos lexicográficos electrónicos multilingües y plurilingües: definición y clasificación tipológico-descriptiva

*María José Domínguez Vázquez*
Universidade de Santiago de Compostela
majo.dominguez@usc.es

**Resumen:** este estudio propone una clasificación de recursos lexicográficos electrónicos multilingües y plurilingües. Por una parte, la aplicación de criterios cuantitativos y cualitativos permitirá llevar a cabo una clasificación tipológica de herramientas lexicográficas, como diccionarios frente a plataformas y portales. Por otra parte, se aborda la distinción de recursos multilingües y plurilingües en términos de mayor o menor prototipicidad. Junto con la descripción de los diferentes recursos, el artículo presenta parámetros y propuestas tipológicas para la definición y delimitación de tipos de recursos electrónicos, en especial el diccionario multilingüe y los portales, y dibuja de manera más nítida el objeto de estudio de la lexicografía multilingüe y la plurilingüe. O, al menos, lo cuestiona.

# Multilingual and plurilingual electronic lexicographical resources: definition and typological-descriptive classification

**Abstract:**The aim of this paper is to provide a classification of multilingual and plurilingual electronic lexicographic resources which would enable, one the one hand, the implementation of quantitative and qualitative criteria to produce a typological taxonomy of lexicographical tools, such as dictionaries, as opposed to platforms and websites and, on the other, the distinction of multilingual and plurilingual resources in terms of their larger or lesser prototyping degree. In addition to offering a thorough description of the different resources, this paper also puts forward some parameters and typological proposals to define and demarcate a number of electronic resources, particularly multilingual dictionaries and portals, while also outlining – and even questioning – the object of study of multilingual and plurilingual lexicography.

# 0. Introducción[1]

En la literatura científica, tanto a los diccionarios multilingües electrónicos[2] como a las características que los confrontan a portales y plataformas no se les ha prestado especial atención desde un punto de vista teórico. Así, algunas definiciones sobre recursos lexicográficos electrónicos suelen obviar la frontera de diccionarios frente a plataformas y portales, quizá porque muchas de ellas se ocupan más bien de recursos impresos que de electrónicos, o quizá porque muchas de ellas son anteriores a la revolución tecnológica actual. Además, se observan solapamientos terminológicos. Así, por ejemplo, los diccionarios multilingües se definen en la obra *Los diccionarios del español en el siglo XXI* (Haensch/Omeñaca, 2004: 59) como un tipo de diccionario plurilingüe, considerándose plurilingües tanto a los bilingües (dos lenguas) como a los multilingües (más de dos). Pirouzan (2009: 18) también recurre al criterio del número de lenguas para categorizar a los diccionarios multilingües. Por su parte, Marello (2003: 336) indica que suelen ser monodireccionales y Porto Dapena clasifica los diccionarios como "monolingües o unilingües, bilingües, y plurilingües o polilingües, también llamados antiguamente políglotas" (Porto Dapena, 2002: 57), con lo cual diferencia –al contrario de Haensch/Omeñaca (2004)– a los bilingües frente a los plurilingües, pero no como una subclase del otro, sino posicionándolos en la clasificación en el mismo nivel jerárquico. Atendiendo a su finalidad, este autor describe los diccionarios plurilingües como recursos para el trabajo técnico o científico ya que aportan terminologías de este tipo (2002: 58). Se observa, por tanto, una definición restrictiva de la lexicografía multilingüe, caracterizada mayoritariamente en relación al número de lenguas y a la tipología de recursos de los que se ocupa, como es el caso de la propuesta de Fuertes-Olivera/Bergenholtz (2019), que acota el ámbito de estudio de la lexicografía multilingüe a un recurso concreto, el diccionario:

> Our above reflections allow us to define multilingual lexicography in the era of the internet as the theory and practice of unified and well-connected monolingual, bilingual and multilingual dictionaries using data from a multilingual database.

---

1. Los resultados de este estudio están en relación con el proyecto "Generador multilingüe de estructuras argumentales del sustantivo con aplicación en la producción en lenguas extranjeras", financiado por FEDER/Ministerio de Economía, Industria y Competitividad – Agencia Estatal de investigación (FFI2017-82454-P), con el proyecto "Generación multilingüe de estructuras argumentales del sustantivo y automatización de extracción de datos sintáctico-semánticos", financiado por la Fundación BBVA(Ayudas Fundación BBVA a Equipos de Investigación Científica 2017), así como con la Red de Lexicografía RELEX (ED341D R2016/046).

2. En las últimas décadas la literatura sobre lexicografía electrónica ha sido muy prolífica (Fuertes- Olivera/ Bergenholtz 2011, Klosa/Müller-Spitzer 2016, Mann 2014, De Schryver, 2003 o Wiegand et al. 2010), si bien menos abundante en propuestas globales, que aúnen aproximaciones lexicográficas mediodependientes, medioindependientes y usuariodependientes en favor de una clasificación de los recursos electrónicos (vid. Domínguez 2017).

> These dictionaries are information tools that cover words, terms, facts, and/or things in several languages, have the same conceptualization at the pre-compilation phase, and make use of lexicographic and technological know-how which allows (a) lexicographers to add new languages to the same information database from which new monolingual, bilingual and multilingual dictionaries can be extracted and (b) users to retrieve connected data easily and to spot and understand possible similarities and differences among the several languages covered.

En definitiva, la silueta de los recursos electrónicos multilingües no está bien dibujada. Consideramos que es imprescindible proporcionar una definición menos generalista –y, por tanto, distintiva– del diccionario multilingüe electrónico, así como una propuesta tipológica más pormenorizada de tipos y clases de recursos electrónicos acorde con sus funcionalidades y usuarios. Para este fin, en 1. se proponen criterios que contribuyen a distinguir tipos de recursos electrónicos multilingües y plurilingües[3]. Como herramientas multilingües clasificamos a aquellas que conceden un mismo tratamiento jeráquico a todas las lenguas descritas, siendo estas siempre más de dos. Por el contrario, el peso descriptivo adjudicado a las lenguas en los recursos plurilingües es diferente (para más detalle, véase 4.). El apartado 2. aborda el diccionario electrónico multilingüe y plurilingüe y propone una categorización en criterios de mayor o menor prototipicidad. De la definición y tipología de portales y plataformas se ocupa el apartado 3. y en el 4. se pone el foco en la lexicografía electrónica multilingüe y plurilingüe.

# 1. Algunos criterios para la delimitación de recursos electrónicos multilingües y plurilingües

Con la finalidad de delimitar los recursos lexicográficos electrónicos multilingües frente a los plurilingües y, a su vez, los diccionarios (vid. 2) frente a portales y plataformas (vid. 3), se proponen los siguientes parámetros cuantitativos y cualitativos:

### Criterios cuantitativos (CCn)

**CCn[1]**: *Número de lenguas*: Un recurso plurilingüe/multilingüe cuenta con un número de lenguas superior a dos.

3. Entre los criterios no se incluyen, por ejemplo, los diferentes soportes desde los que el usuario puede acceder al recurso o herramienta, tales como tabletas o teléfonos móviles, puesto que el objetivo de este estudio es la delimitación de las herramientas plurilingües frente a las multilingües y el desglose de características de los diccionarios frente a los portales. En esta contribución se ejemplifica con diccionarios accesibles en red.

**CCn**[2]: *Metalengua(s)*: Este criterio se relaciona con el anterior, ya que en un recurso prototípico ideal el número de metalenguas tendría que ser parejo al número de lenguas analizadas.

**Criterios cualitativos (CCl)**

**CCl**[1]: *Estatus de las lenguas y grado de profundidad de análisis*: Una herramienta multilingüe paradigmática concede a todas las lenguas el mismo estatus jerárquico en cuanto a la cantidad y calidad de la información aportada sobre las mismas. En caso contrario, se trata de un recurso plurilingüe.

**CCl**[2]: *Interconexión de datos y tipos de búsqueda*: Atendiendo a criterios como la comparabilidad y equivalencia de las lenguas, son objeto de análisis:

- la interconexión y la hipervinculación de los módulos de información de los subdiccionarios que conforman la herramienta multilingüe/plurilingüe.

- las opciones de acceso y búsqueda de datos, como 1>n, 1<n y n<>n[4]. La permanencia en la interfaz de consulta en una búsqueda concreta también se antoja relevante para el análisis.

Por tanto, no se trata solo de señalar si un recurso permite una consulta monolingüe, bilingüe o multilingüe/plurilingüe, sino, además, de observar en qué medida esto afecta a conceptos clave como la lengua de partida y destino de un diccionario, así como de poner en valor conceptos como la *multilateralidad* y la *consulta cross-lingual*.

**CCl**[3]: *Índice interlingüístico y datos relacionales*: En los recursos que abordan más de dos lenguas, la decisión de qué servirá de nexo de unión entre ellas, por ejemplo vocablos o formas frente a conceptos o acepciones de significado, resulta central. Este aspecto no afecta únicamente al tipo de recurso y su macroestructura, sino que también supone decisiones relativas al diseño de la base de datos, tales como el fichado y almacenamiento, las futuras opciones de búsqueda o el enlace de los datos de las diferentes lenguas contempladas, entre otros.

**CCl**[4]: *Tipos de acceso a los datos*: Siguiendo a Engelberg/Müller-Spitzer (2013), contemplaremos 3 estructuras de acceso:

4. 1>n se refiere a la relación 'una lengua con todas', 1<n contempla la relación 'todas las lenguas con una' y n<>n se refiere a la relación 'todas con todas'.

- *External access structures* (externas): acceso a la interfaz externa del diccionario;
- *Outer access structures* (exteriores): acceso al diccionario mediante una búsqueda por lemas de los subdiccionarios o diccionarios integrados;
- *Inner access structures* (internas): acceso a la información desde diferentes bloques de la microestructura de las entradas de los subdiccionarios o diccionarios integrados.

**CCI[5]:** *Grado de autonomía de subdiccionarios o recursos integrados en una herramienta*: Esta cuestión afecta a la concepción inicial de los subdiccionarios o recursos integrados como recursos individuales o no, y, por tanto, a su consulta como herramientas independientes o no.

**CCI[6]:** *Integración de recursos y datos*: La integración entre recursos y datos es factible cuando los datos o recursos son de un mismo propietario o este tiene derechos prioritarios sobre ellos. Se observa este aspecto, por ejemplo, cuando entre ellos hay referencias cruzadas.

**CCI[7]:** *Diseño del recurso y sus componentes*: Estrechamente ligado a CCI[5] y CCI[6], contempla si el diseño de los subdiccionarios o recursos integrados es uniforme.

# 2. Diccionarios electrónicos multilingües y plurilingües

## 2.1 Introducción

Qué es un diccionario electrónico multilingüe o plurilingüe y qué los caracteriza y los diferencia frente a otros recursos es la cuestión central que se plantea aquí. Para intentar dar respuesta a esta cuestión recurrimos, en primer lugar, a la definición de *diccionario electrónico* de Müller-Spitzer (2008: 50). Partiendo de esta, Domínguez (2017:183) define el diccionario multilingüe electrónico como "una obra de consulta electrónica cuya finalidad genuina consiste en que un potencial usuario pueda conseguir a través de datos lexicográficos información multilingüe sobre un objeto lingüístico". Sin embargo, esta definición resulta insuficiente, puesto que, además, un diccionario multilingüe "is not a collection of different dictionaries juxtaposed, but the same dictionary with the same elements for all the languages covered" (Fuertes-Olivera/Bergenholtz, 2019)[5], criterio por el cual se diferencia de portales y plataformas (vid. 3). Desgranamos, a continuación, otros criterios evaluadores.

## 2.2 Lenguas y conceptualización de la obra

El criterio cuantitativo relativo al **número de lenguas** (vid. 1, CCn$^{1}$), tan comúnmente manejado, sirve para confrontar recursos monolingües frente a multilingües y plurilingües –siempre más de dos– pero no es de aplicación en la diferenciación de estos dos últimos. Entendemos que a este criterio cuantitativo hay que sumarle uno cualitativo: el estatus y tratamiento de las lenguas en los recursos. Así, por ejemplo, la consulta de 'paraula' en el *Diccionari de la Llengua Catalana multilingüe*[6] arroja datos sobre las acepciones en catalán y una serie de equivalencias: catalán>español, catalán>inglés, catalán>alemán y catalán>francés (Imagen 1). Por el contrario, las búsquedas bilaterales nos conducen desde cualquiera de estas lenguas al catalán, pero no a cualesquiera de las otras lenguas (Imágenes 2 y 3)[7].

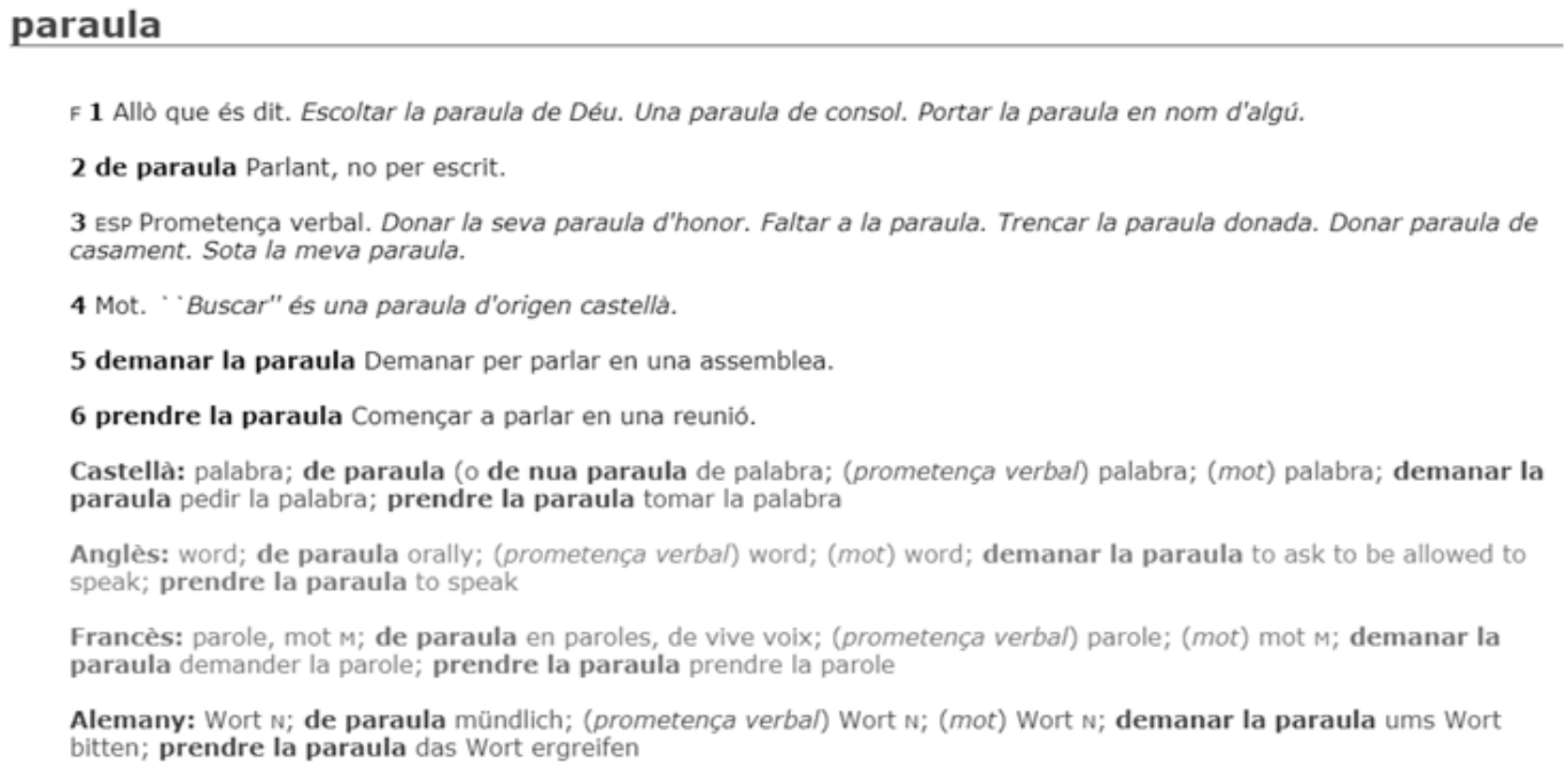
**paraula**

F **1** Allò que és dit. *Escoltar la paraula de Déu. Una paraula de consol. Portar la paraula en nom d'algú.*

**2 de paraula** Parlant, no per escrit.

**3** ESP Prometença verbal. *Donar la seva paraula d'honor. Faltar a la paraula. Trencar la paraula donada. Donar paraula de casament. Sota la meva paraula.*

**4** Mot. *``Buscar'' és una paraula d'origen castellà.*

**5 demanar la paraula** Demanar per parlar en una assemblea.

**6 prendre la paraula** Començar a parlar en una reunió.

**Castellà:** palabra; **de paraula** (o **de nua paraula** de palabra; (*prometença verbal*) palabra; (*mot*) palabra; **demanar la paraula** pedir la palabra; **prendre la paraula** tomar la palabra

**Anglès:** word; **de paraula** orally; (*prometença verbal*) word; (*mot*) word; **demanar la paraula** to ask to be allowed to speak; **prendre la paraula** to speak

**Francès:** parole, mot M; **de paraula** en paroles, de vive voix; (*prometença verbal*) parole; (*mot*) mot M; **demanar la paraula** demander la parole; **prendre la paraula** prendre la parole

**Alemany:** Wort N; **de paraula** mündlich; (*prometença verbal*) Wort N; (*mot*) Wort N; **demanar la paraula** ums Wort bitten; **prendre la paraula** das Wort ergreifen

Imagen 1: 'paraula' en el *Diccionari de la Llengua Catalana multilingüe*

5. Véase también en la sección 1 los criterios CCl$^{5}$, CCl$^{6}$ y CCl$^{7}$, por ejemplo.

6. Las referencias completas de los recursos citados a lo largo de la obra se encuentran en la bibliografía.

7. Esta microestructura también se observa, por ejemplo, en DiGaTIC – *Dicionario galego das TIC*, así como en glosarios multilingües (Domínguez/Mirazo 2016).

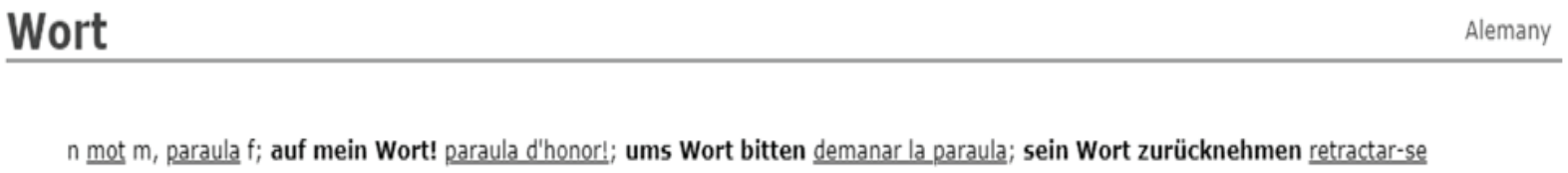
**Wort** Alemany

n mot m, paraula f; **auf mein Wort!** paraula d'honor!; **ums Wort bitten** demanar la paraula; **sein Wort zurücknehmen** retractar-se

Imagen 2: 'Wort' en el *Diccionari de la Llengua Catalana multilingüe*

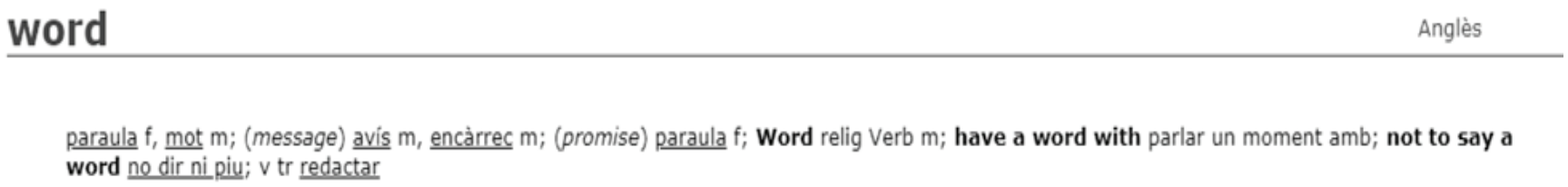
**word** Anglès

paraula f, mot m; (*message*) avís m, encàrrec m; (*promise*) paraula f; **Word** relig Verb m; **have a word with** parlar un moment amb; **not to say a word** no dir ni piu; v tr redactar

Imagen 3: 'word' en el *Diccionari de la Llengua Catalana multilingüe*

Esto supone, por tanto, que desde determinadas lenguas la única búsqueda reversible posible es bilingüe y fijada a pares determinados (no multilingüe, no *cross-lingual*) y, por tanto, consideramos este recurso plurilingüe (siguiendo un criterio cuantitativo), pero no multilingüe (según un parámetro cualitativo), ya que el peso concedido a las lenguas es diferente. Asimismo, bajo el paraguas de los recursos plurilingües, se observan diferencias notables relativas a la información aportada sobre las lenguas descritas. Así, el *Diccionario electrónico multilingüe de verbos de movimiento* analiza un número mayor de lenguas que el *Diccionari de la Llengua Catalana multilingüe*, pero la información aportada sobre estas está reducida a determinados módulos, como, por ejemplo, la traducción de ejemplos en el primero de ellos.

Bajo el epígrafe de multilingüe se observa la misma casuística en cuanto a criterios de mayor o menor prototipicidad. En esta línea, *The Internet Picture Dictionary*[8] (Imagen 4) puede ser considerado un diccionario multilingüe, pero no prototípico en sentido ideal, puesto que los subdiccionarios no están interconectados y no permiten búsquedas comparativas, presentándose la información de modo monolingüe. Por su parte, MobiLex (Gouws, 2019), diccionario en formato app, ofrece información de determinados campos para el Afrikaans, Inglés y Xhosa y representa un modelo lexicográfico multilingüe diferente al anterior, puesto que, por ejemplo, podemos obtener información de las tres lenguas de modo simultáneo.

8. El *Internet Picture Dictionary* es un diccionario de imágenes multilingüe en línea diseñado especialmente para estudiantes de estudiantes de segunda lengua y principiantes de inglés, francés, alemán, español e italiano de todas las edades.

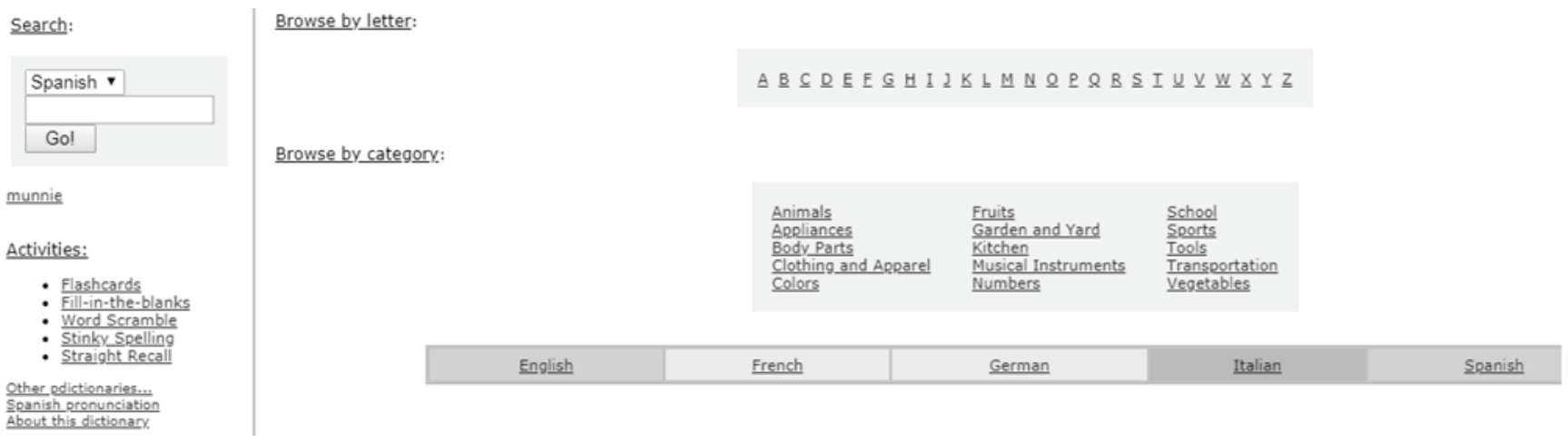

Imagen 4: *The Internet Picture Dictionary: Spanish*

Más representativo del tipo diccionario multilingüe es el diccionario PORTLEX (Domínguez/Valcárcel, 2019), el cual ha sido concebido desde un principio como un recurso que ofrece los mismos campos informativos para las 5 lenguas contempladas[9] (Imagen 6). Si comparamos este recurso con los *Wiktionaries*, concluiremos que existen diferencias notables relativas al tratamiento cuantitativo y cualitativo de las lenguas y al carácter multilingüe o plurilingüe de ambos recursos, lo cual nos ayuda a perfilar nuestra propuesta tipológica. A diferencia de PORTLEX, y a pesar de su nombre, el *Wiktionary* no es un diccionario multilingüe en sentido estricto, sino una plataforma de diccionarios colaborativos abiertos plurilingües, la cual alberga *sites* particulares para cada una de sus ediciones, en donde, por tanto, hay un diccionario plurilingüe en español, otro en francés, otro en alemán, etc. (vid Valcárcel 2017: 383-384; Imagen 5). El *Wiktionary* contiene, por tanto, diccionarios monolingües con una interfaz común[10].

En cuanto a la **metalengua**, un diccionario multilingüe ideal debería contemplar tantas metalenguas como lenguas descritas[11]. La presencia de diferentes metalenguas en un recurso supondría que pudiéramos considerar, desde este punto de vista, una bidireccionalidad o multidireccionalidad plena.

## 2.3 Índice interlingüístico y datos relacionales

Muchos de los recursos multilingües actuales se conforman con diccionarios monolingües/bilingües ya existentes, a lo que se debe añadir que los segundos no son siempre completamente bilaterales. Esto supone un primer obstáculo a la hora de enlazar los datos relativos a las lenguas y da lugar a que muchos recursos –que se denominan a sí mismos multilingües– sean *de facto* un conjunto o compendio de diccionarios bilingües. Además, la dificultad de elaborar un diccio-

9. PORTLEX puede definirse como un diccionario multilingüe *cross-lingua* (Domínguez 2017: 190-196). Esta herramienta permite una consulta monolingüe, bilingüe y multilateral *cross-lingua* para 5 lenguas: alemán, español, francés, gallego e italiano.

10. Cabe señalar que su objetivo no es presentar información lexicográfica de lenguas en contraste.

11. En un caso dado, este aspecto ejemplificaría claramente las ventajas de la obra electrónica frente a la limitación física de la obra impresa.

nario multilingüe va en aumento, si tenemos en cuenta que los recursos bilingües cuentan, por norma general, con entradas asociadas a formas o vocablos y no a conceptos (vid. 1, CCl3).

Una vinculación de datos siguiendo formas (*wordforms*) está presente en el *Wiktionary* (vid. 2.2.). Esto se observa en las entradas de una edición concreta en la que se incluyen homógrafos de otras lenguas:

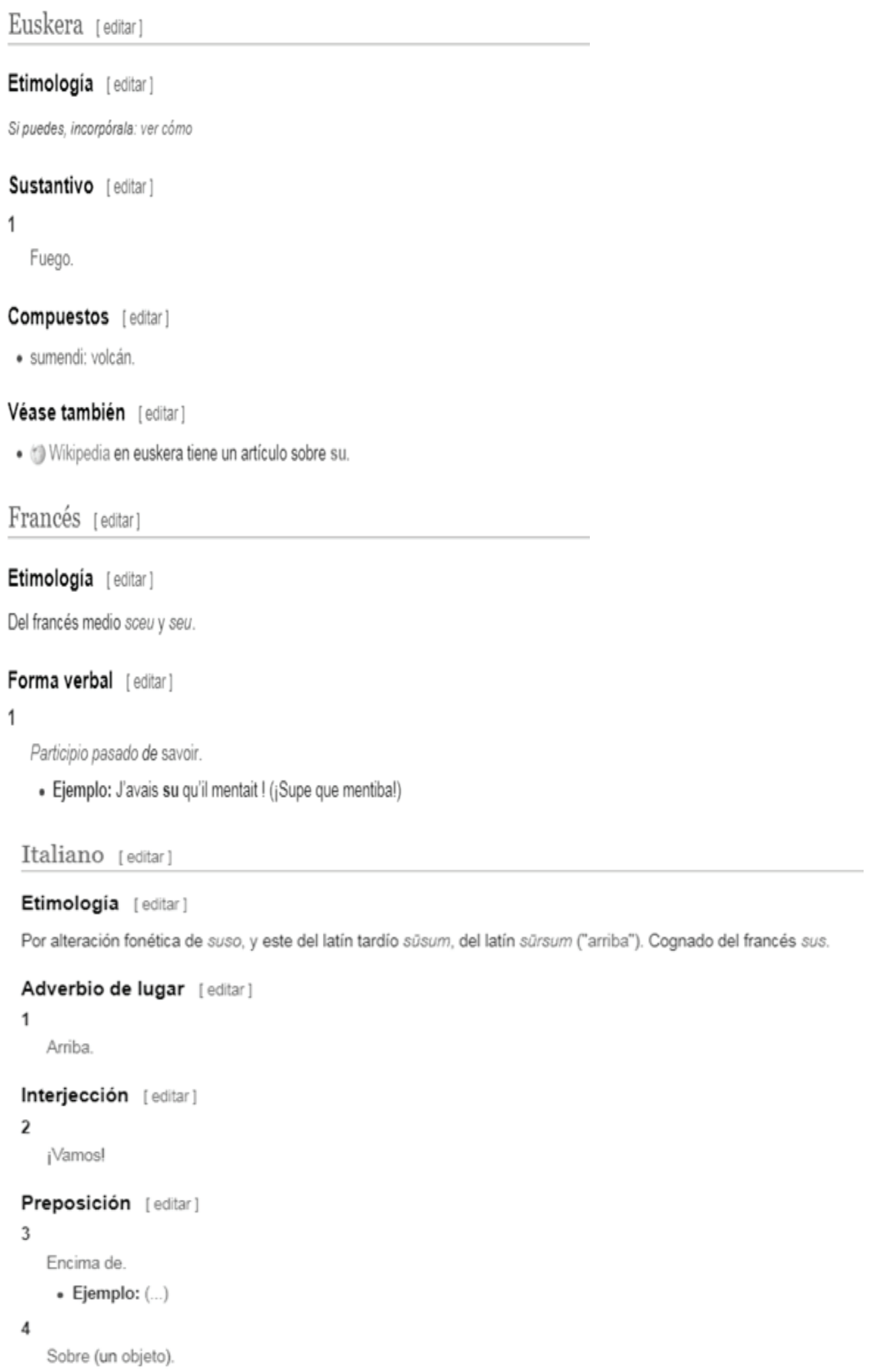

Euskera [ editar ]

**Etimología** [ editar ]

*Si puedes, incorpórala: ver cómo*

**Sustantivo** [ editar ]

1
Fuego.

**Compuestos** [ editar ]

- sumendi: volcán.

**Véase también** [ editar ]

- Wikipedia en euskera tiene un artículo sobre su.

Francés [ editar ]

**Etimología** [ editar ]

Del francés medio *sceu* y *seu*.

**Forma verbal** [ editar ]

1
*Participio pasado* de savoir.

- **Ejemplo:** J'avais **su** qu'il mentait ! (¡Supe que mentiba!)

Italiano [ editar ]

**Etimología** [ editar ]

Por alteración fonética de *suso*, y este del latín tardío *sūsum*, del latín *sūrsum* ("arriba"). Cognado del francés *sus*.

**Adverbio de lugar** [ editar ]

1
Arriba.

**Interjección** [ editar ]

2
¡Vamos!

**Preposición** [ editar ]

3
Encima de.

- **Ejemplo:** (...)

4
Sobre (un objeto).

Imagen 5: Búsqueda 'su' en el *Wiktionary*: Presencia de vocablos de otras lenguas[12]

12. La información aquí aportada es una selección de la entrada.

Por el contrario, obras como los *Diccionarios de contabilidad* (Fuertes-Olivera, en prensa), *MobiLex* (Gouws, 2019) o PORTLEX recurren a conceptos o significados comunes, que sirven de índice (Imagen 7). En definitiva, en la elaboración de un diccionario multilingüe, en sentido ideal, el índice interlingüístico no pueden ser las *palabras* o formas léxicas, sino los conceptos (vid. en 1, CCl[3]), tal y como se denominan en *OmegaWiki* (vid. 3), o las acepciones, como en el caso de PORTLEX[13]. Para tal finalidad se requiere un diseño específico de la relación en la consulta de los datos, como es el caso de los dos últimos recursos citados. Así, en el diccionario PORTLEX, cada idioma cuenta con un módulo específico en el que se almacenan los datos relativos a cada uno de ellos. De este modo, como se observa en la imagen 6, para cada módulo (en este caso la combinatoria actancial con las preposiciones *de+sobre* en español del sustantivo 'texto') se obtiene la misma información alineada y contrastiva en cualquiera de las lenguas contempladas en el recurso.

| Realización formal | Rasgo categorial | Tipo complemento |
| --- | --- | --- |
| de | Humano, Institución | Complemento sujeto |
| sobre | Inmaterial | Complemento prepositivo |

**Frase tipo:** El texto del poeta sobre el amor

El volumen incluye la reproducción del atlas original y el manuscrito con **el texto** del autor sobre el contenido de los mapas y la historia de su ejecución.
CREA: El País: El descubrimiento del atlas de Texeira cambia la cartografía del siglo XVII, Diario El País, S.A.: Madrid, 30/10/2002.

| Realización formal | Rasgo categorial | Tipo complemento |
| --- | --- | --- |
| Genitivo | Humano, Institución | Complemento sujeto |
| über + Acusativo | [–]: sin restricciones | Complemento prepositivo |

**Frase tipo:** Der Text des Philosophen über die Schönheit

In der ersten Ausgabe ist es ein **Text** des Philosophen Hans-Georg Gadamer über die Schönheit mit einem Kommentar von Odo Marquard.
DEREKO: BRZ07/MAR.15990 Braunschweiger Zeitung, 10.03.2007; „Alte Hüte" für den Untergrund.

| Realización formal | Rasgo categorial | Tipo complemento |
| --- | --- | --- |
| von + Dativo | Humano, Institución | Complemento sujeto |
| über + Acusativo | [–]: sin restricciones | Complemento prepositivo |

**Frase tipo:** Der Text von Peter über seine Weltreise

Im Furtwänglerpark gegenüber dem Festspielhaus präsentiert der Franzose Jeanluc Ducourt seine Solo- Performance und Soundinstallation zu einem **Text** von Robert Anthelme über seine Gefangenschaft in Buchenwald.
DEREKO: N00/JUL.31780 Salzburger Nachrichten, 12.07.2000, Ressort: LOKALES; Ein Fest für alle.

| Realización formal | Rasgo categorial | Tipo complemento |
| --- | --- | --- |
| Compuesto | Humano, Institución | Complemento sujeto |
| über + Acusativo | [–]: sin restricciones | Complemento prepositivo |

**Frase tipo:** Der Regierungstext über die neuen Gesetze

Gestern hat die zuständige Parlamentskommission den Bericht zu dem Regierungs**text** über den Zugang zur Palliativbehandlung angenommen.
DEREKO: LTB08/FEB.00409 Luxemburger Tageblatt, 15.02.2008; Letzte Reflexionselemente vor dem Votum.

Imagen 6: Datos alineados contrastivos en *PORTLEX*

---

13. En este sentido, PORTLEX opera de modo parejo a herramientas multilingües basadas en *WordNet*, que se organiza en sentidos semánticos. Cada uno de estos sentidos está lexicalizado en un conjunto de sinónimos, denominado "synset". En el modelo EuroWordNet son los sentidos los que actúan como "Interlingual Index", esto es, como índices interlingüísticos (Solla/Gómez 2015: 171–172). *WordNet* ha sido denominada por Fellbaum (1998), en algunas ocasiones, diccionario semántico.

## 2.4 Grado de autonomía, integración de los subdiccionarios, acceso y diseño[14]

Para salvar alguno de los obstáculos señalados previamente, en el desarrollo de diccionarios multilingües se viene aplicando el *hub-spoken-model* o el *polytypological mother dictionary* (Gouws, 2014)[15], como es el caso del diccionario PORTLEX, en el que "the mother dictionary was constituted by a list of meanings extracted for the lexemes analyzed in the Spanish language, which also provides the metalanguage of the dictionary" (Domínguez/Valcárcel, 2019). Estos significados o definiciones forman las claves de la relación interlingüística dentro de la base de datos y también sirven para seleccionar las equivalencias que se analizarán en los diferentes idiomas[16] (vid. 2.2.):

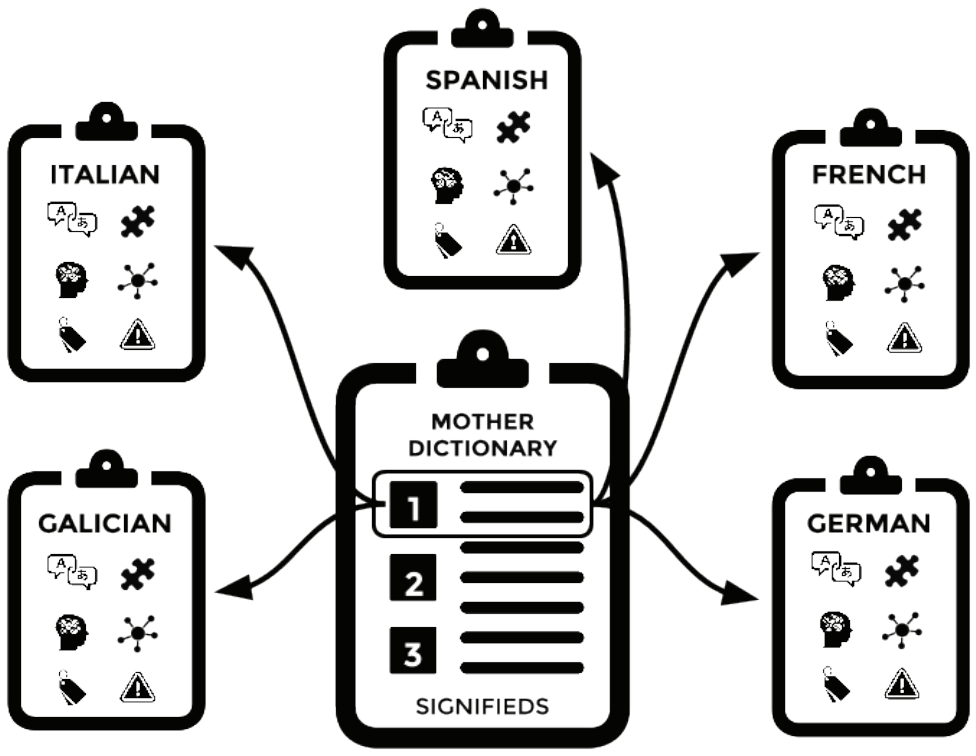

Imagen 7: Diccionario madre e índice relacional en *PORTLEX*

14. Estos parámetros CCl[4]-CCL[7] han sido propuestos por Engelberg/Müller-Spitzer (2013) en su clasificación de portales lexicográficos. Es de interés, a continuación, sopesar si pueden ser también de aplicación –mediante una reformulación– para la clasificación de diccionarios multilingües/plurilingües o incluso para la delimitación de estos frente a las plataformas y portales.

15. Siguiendo a Gouws, los elementos léxicos de las lenguas incluidas se vinculan mediante una lengua central común.

16. Entre otros criterios, para justificar la selección de un equivalente en una lengua determinada, dicho equivalente debe tener un significado parejo en los respectivos diccionarios de referencia de dicha lengua. Sin embargo, más allá de sus limitaciones (Mirazo 2016: 99-101), operar con el concepto de diccionario madre y adoptar el significado como una clave relacional ha hecho posible el diseño multilingüe que propone PORTLEX (Domínguez /Valcárcel, 2019).

Ya que cada lengua constituye dentro del diccionario un producto independiente (si bien con ciertas restricciones, véase nota al pie 15), los subdiccionarios de los diccionarios multilingües del tipo de PORTLEX muestran una alta autonomía.

El diccionario multilingüe, en sentido ideal, tendría que contemplar las tres estructuras de acceso a los datos señaladas en CCl[4] (esto es, externa, exterior e interna), asegurando esta última la relación multilateral y *cross-lingual*[17]. Por tanto, los subdiccionarios deberían estar altamente interconectados e integrados. En definitiva, lo realmente definitorio sería que todos los módulos informativos estuvieran enlazados entre sí y que existiesen opciones de búsqueda desde todas las lenguas y campos informativos. En cuanto a su diseño, un diccionario multilingüe prototípico tiene que estar altamente unificado, lo cual suele ser habitual, si el recurso ha sido concebido desde su origen como tal.

## 2.5 Hacia el diccionario multilingüe prototípico

Para caracterizar al diccionario multilingüe prototípico se puede operar mediante una escala de mayor o menor prototipicidad, en la que las siguientes categorías constituyen el mínimo definitorio[18]:

**Tabla 1: Diccionario multilingüe según grados de prototipicidad**

| | | | | | |
|---|---|---|---|---|---|
| **Relación lingüística** | Lengua$^1$ >Lengua$^n$ (una con todas) | Lengua$^1$ <Lengua$^n$ (todas con 1) | | Lengua$^n$ ><Lengua$^n$ (todas todas) | |
| **Metalengua** | Metalengua$^1$ #Lengua$^n$ | | | Metalengua$^n$ = Lengua$^n$ | |
| **Vinculación entre contenidos de subdiccionarios** | Lema | Módulos$^x$ | | Módulos$^n$ | |
| **Acceso a la información** | monolingüe | bilingüe | multilingüe | multilingüe multilateral | multilingüe *cross-lingual* |
| | − | | | | + |

17. En algunos diccionarios multilingües se observa que la relación de equivalencia se centra en los lemas de las diferentes lenguas, con lo cual el diccionario multilingüe se puede acercar en estos casos estructuralmente al glosario.

18. Bajo *relación lingüística* se plasma la interconexión de datos y tipos de consulta. Se atiende aquí a conexiones entre lenguas, tales como "desde una lengua se accede a todas" (Lengua$^1$>Lengua$^n$), "todas las lenguas refieren a una" (Lengua$^1$<Lengua$^n$), o "desde todas las lenguas se accede a todas" (Lengua$^n$<>Lengua$^n$). En cuanto a la *metalengua*, la tabla refiere al uso de una única metalengua para todas las lenguas descritas en el recurso (Metalengua$^1$#Lengua$^n$) o a la presencia de tantas metalenguas como lenguas analizadas (Metalengua$^n$=Lengua$^n$). La *vinculación entre contenidos* ejemplifica si esta se ciñe únicamente al lema, a determinados módulos o a todos los módulos del recurso. El último bloque presenta diferentes *accesos* a la información. Para más detalle, véase el apartado 1.

# 3. Portales y plataformas multilingües y plurilingües

## 3.1. Definición y tipos

Es complejo establecer una estricta delimitación entre diferentes tipos de portales y plataformas, ya que contamos con portales que comparten características de más de un tipo (Boelhouwer/Dykstra/Sijens, 2017: 759). No se trata, sin embargo, de un fenómeno que afecte exclusivamente a la tipología de los portales, ya que la lexicografía electrónica está sufriendo un fenómeno de hibridización (Granger, 2012:4).

Los portales lexicográficos –una página de internet o un conjunto de páginas interconectadas, que permiten el acceso a diferentes diccionarios en internet o a los datos que estos contienen (Engelberg/Müller-Spitzer, 2013)– se caracterizan por tres aspectos centrales:

> i) that is presented as a page or set of interlinked pages on a computer screen and
>
> ii) provides access to a set of electronic dictionaries
>
> iii) where these dictionaries can also be consulted as standalone products.
>
> (Engelberg/Müller-Spitzer, 2013: 1024)

Engelberg/Müller-Spitzer (2013) y Engelberg/Storrer (2016) proponen para los portales lexicográficos las siguientes clases: repositorios de recursos lexicográficos (*dictionary collection*), agregadores de diccionarios (*dictionary search engine*) y redes de diccionarios (*dictionary net*), a los que cabe añadir los portales virtuales. Se presentan, a continuación, algunos ejemplos:

- **Repositorios de recursos lexicográficos** (*dictionary collection*)[19]

Se trata de compendios de diccionarios, que no tienen por qué contar con interconexión ni integración entre sí, y que pertenecen a autores o gestores de contenido diferentes a los del repositorio. Por este motivo, en la mayor parte de los casos, el acceso a los diccionarios supone abandonar la página de consulta. Sirve de ejemplo el apartado sobre diccionarios del periódico *El Mundo*:

19. En cuanto al número de diccionarios integrados en el portal, Boelhouwer/Dykstra/Sijens (2017) indican que aquellos portales que incluyen más de 30 diccionarios son generalmente repositorios o agregadores.

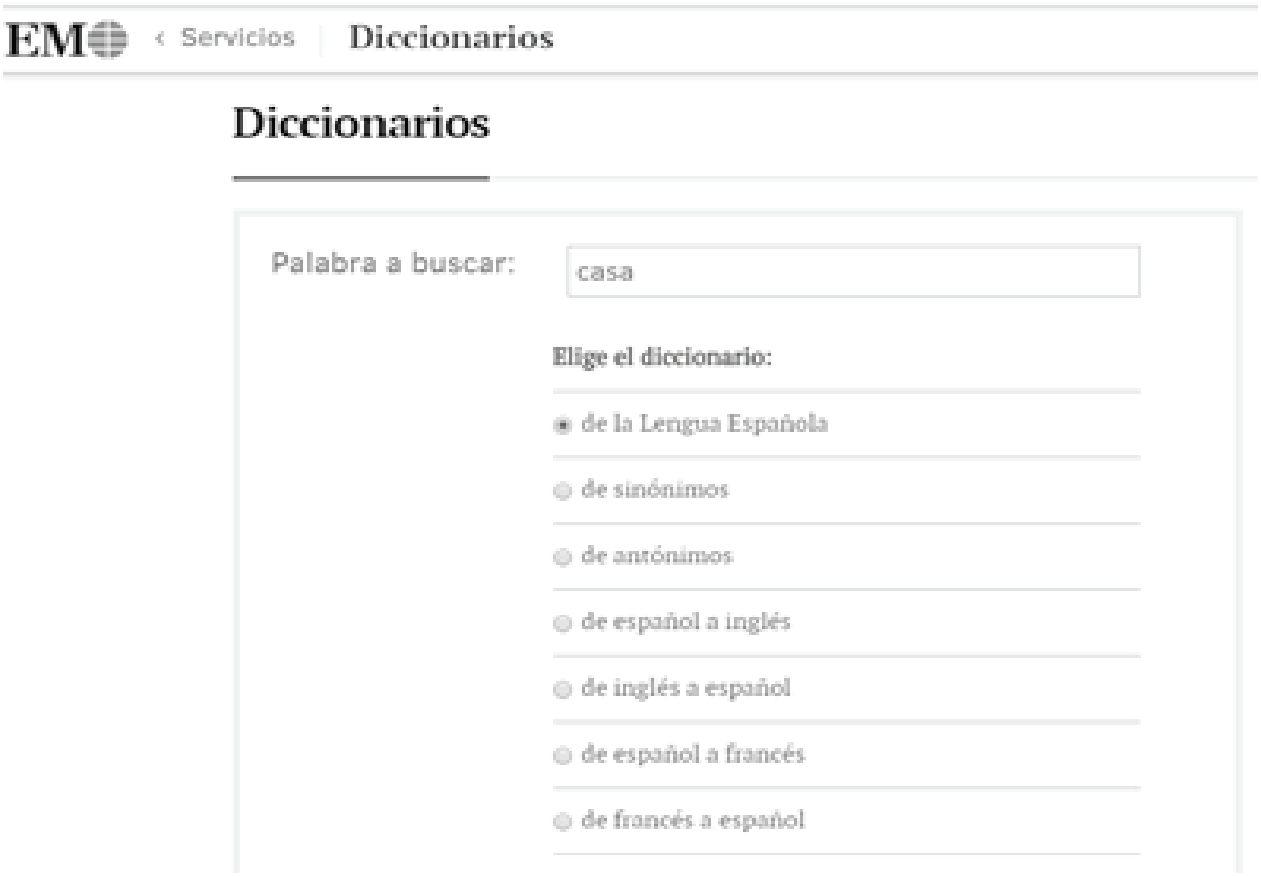

Imagen 8: Diccionarios en *El Mundo*

Más excepcionales son los repositorios con una interfaz de búsqueda integrada: aquellos que desde la página inicial de consulta posibilitan el acceso a algunas de las herramientas compendiadas, como es el caso de la *Erlanger Liste*:

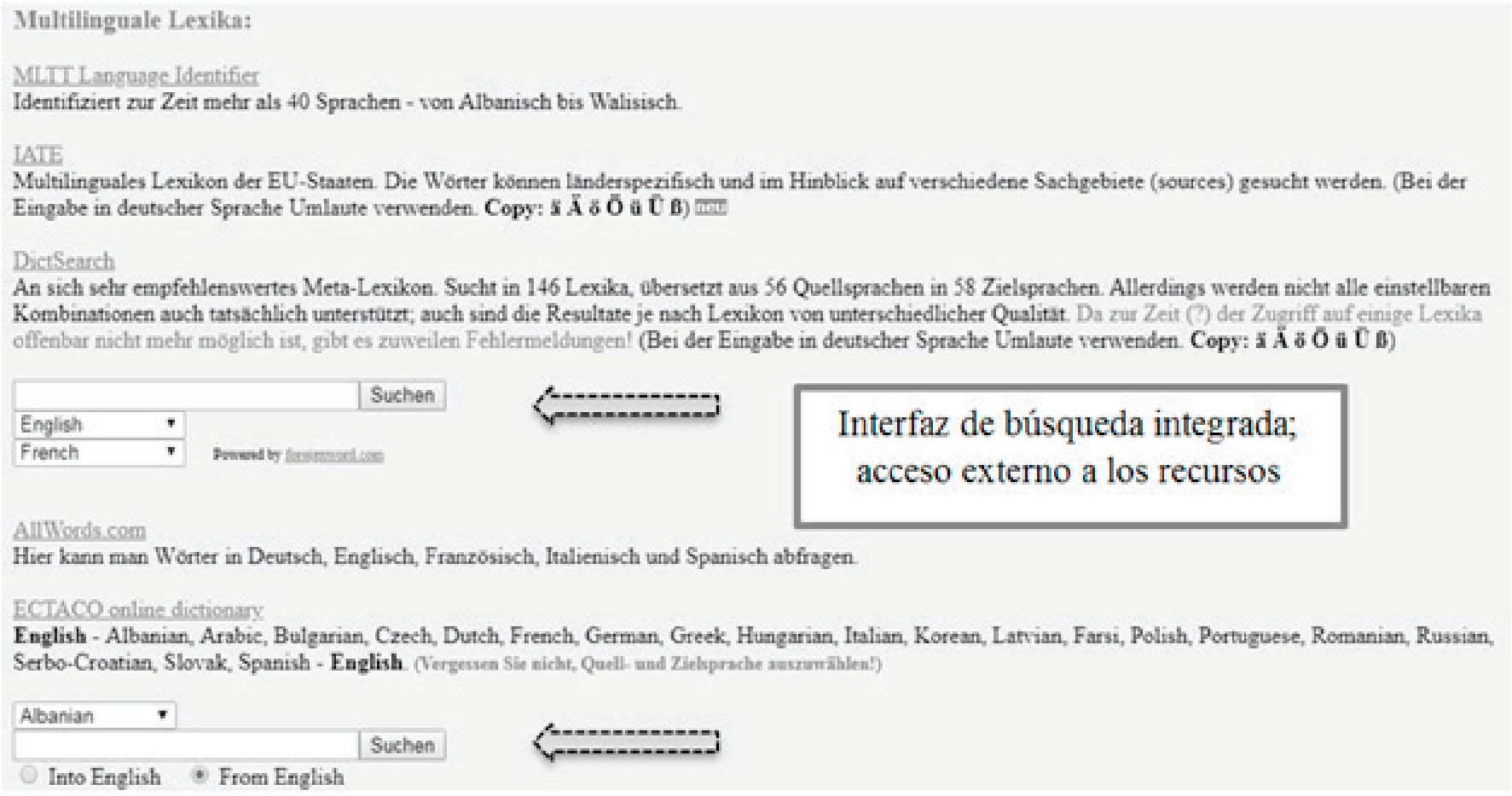

Imagen 9: Ejemplo de interfaz de búsqueda integrada

- **Agregadores de diccionarios** (*dictionary search engine*)

Los metadiccionarios o motores de búsqueda seleccionan información en otros recursos (frecuentemente diccionarios bilingües sin interconexión fuera del par de lenguas). Bajo el paraguas de metadiccionarios se incluyen herramientas que muestran diferencias notables, tanto en las opciones de búsqueda –teniéndose que abandonar o no la interfaz de consulta– como en el tipo de información que aportan. Ejemplos de este tipo de portal son *OneLook*[20], *Lexikologos*[21] o *The free dictionary*[22].

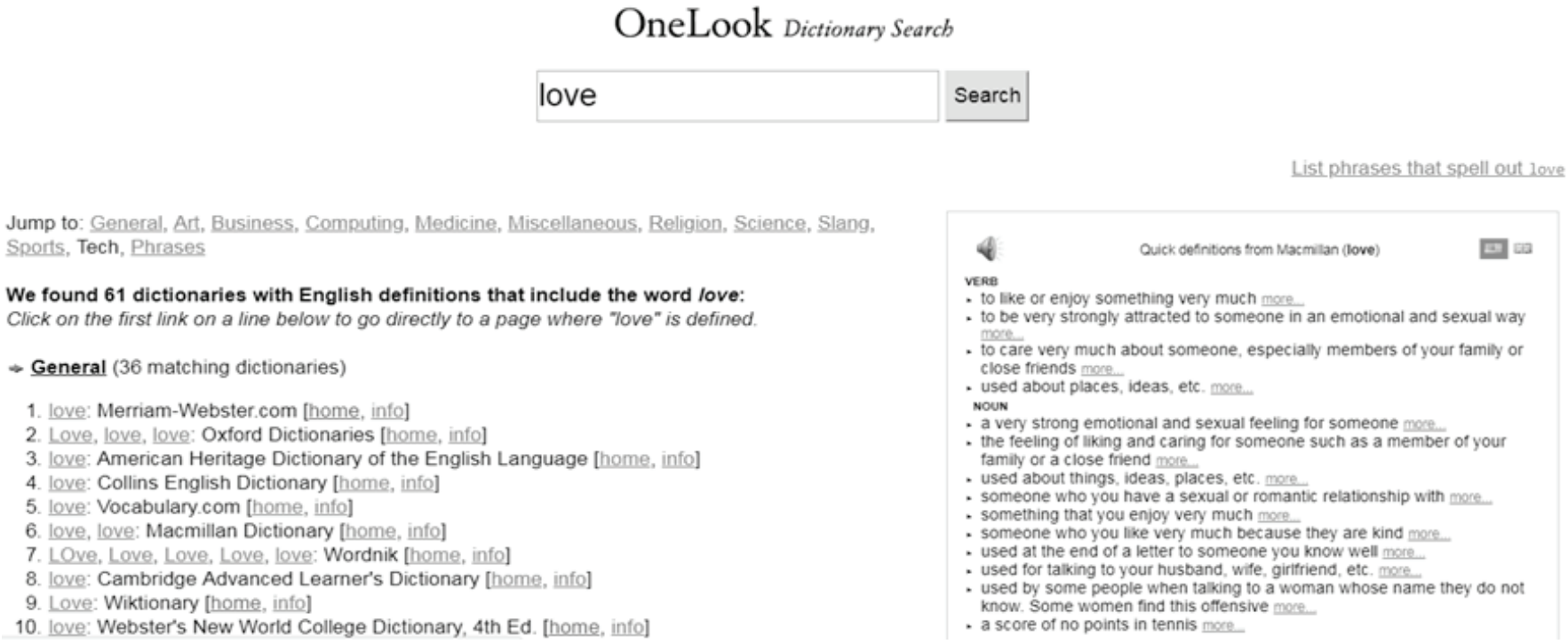

Imagen 10: *OneLook* como ejemplo de agregador de diccionarios

- **Redes de diccionarios** (*dictionary net*)

En las redes de diccionarios, el propietario del portal suele serlo también de los contenidos digitales –o, cuando menos, tiene un acceso privilegiado a ellos (Engelberg/Müller-Spitzer, 2013: 1030). Esto facilita la indexación entre los diccionarios y datos integrados. El conjunto de diccionarios que ofrece la *Real Academia Española* ejemplifica este tipo de recursos (Imagen 11), así como el portal OWID y el de *Larousse*, con obras de referencia de Larousse y Vox.

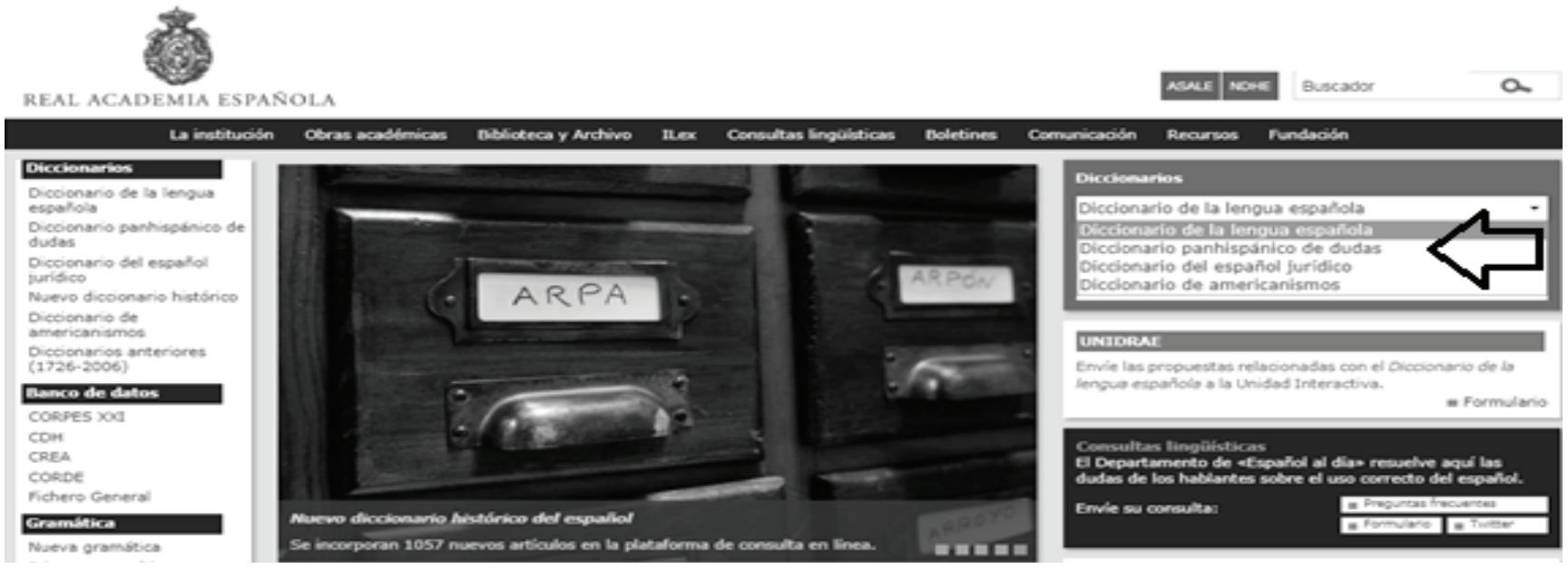

Imagen 11: Portal de la Real Academia Española. Red de diccionarios

20. Atendiendo a diferentes recursos y registros presenta el lema seleccionado en una búsqueda sin filtros. Indexa más de 1.000 diccionarios. En el *OneLookThesaurus* la tipología de búsquedas es muy variada y es posible hacer consultas onomasiológicas, conceptos relacionados e incluso obtener respuestas a preguntas.

21. Es preciso indicar la palabra consultada y seleccionar el recurso de entre la paleta que se ofrece.

22. Presenta el lema dotándolo de información (definición, clase de palabra, etc.), la fuente y, al final, la traducción al idioma seleccionado. Tanto *OneLook* como *the free dictionary* (<http://www.thefreedictionary.com/_/help/help2.htm#22 111>) permiten personalizar los datos así como el sitio web.

- **Portales virtuales**

En los portales virtuales la integración de los recursos y su interconexión es muy elevada, existe una clara homogeneización del diseño y de las estructuras de acceso, la autonomía de los diccionarios indexados es muy baja y casi no se les reconoce como productos independientes (Engelberg/Storrer, 2016). Un ejemplo lo constituyen portales como *Leo*, *Linguee*[23] o *dictionarist*.

En definitiva, a partir de la literatura científica y del análisis de los diferentes recursos se pueden compendiar las características centrales de los portales como sigue:

**Tabla 2: Tipos de portales en contraste**

| | **Repositorios** (*dictionary collection*) | **Agregadores** (*dictionary search engine*) | **Redes** (*dictionary net*) | **Portales virtuales** |
|---|---|---|---|---|
| **Autonomía de los diccionarios** | muy alto grado de autonomía | muy alto grado de autonomía | alto grado de autonomía | baja autonomía |
| | + | | | − |
| **Integración de los recursos** | sin integración | baja integración | alto grado de integración | muy alto grado de integración |
| | − | | | + |
| **Acceso a los recursos** | acceso externo al enlace de los diccionarios | acceso externo a la lista de lemas de los diccionarios, acceso interno | alto grado en diferentes accesos, estructuras de acceso más elaboradas | altamente interconectados |
| | − | | | + |
| **Referencias cruzadas** | recursos totalmente independientes, sin referencias cruzadas entre los diccionarios | totalmente independientes, sin referencias cruzadas entre los diccionarios | sin intervención en los diccionarios, con referencias cruzadas entre los diccionarios | intervención en los diccionarios, casi no se reconoce la autonomía |
| | − | | | + |
| **Diseño** | sin diseño uniforme | sin diseño uniforme | diseño unificado del portal y los diccionarios | unificación en estructuras de acceso y diseño |
| | − | | | + |

Muestran ciertas similitudes con algunos tipos de portales (como, por ejemplo, con las redes de diccionarios) **las plataformas de recursos integrados**, puesto que el propietario del portal también lo es de los recursos integrados o tiene derechos preferentes sobre ellos. La principal diferencia radica, a nuestro parecer,

23. *Linguee* ofrece un gran corpus de traducciones a modo de textos paralelos.

en que en una plataforma pueden estar indexados otros tipos de recursos que no sean estrictamente diccionarios y que los recursos contenidos en las plataformas no tienen por qué haber pasado por ningún filtro lexicográfico. En cuanto a las plataformas de recursos sirven de ejemplo *Babelnet*[24] o la plataforma de *Recursos integrados da Lingua Galega*-RILG.

Abordaremos a continuación otros criterios manejados en el apartado 1, si bien de modo general puesto que no es posible un tratamiento individual y específico para cada uno de los tipos de portales señalados.

## 3.2 Lenguas y conceptualización de la obra

El papel que desempeñan las lenguas –en cuanto a su número y en cuanto al estatus y grado de profundidad de análisis– se puede abordar por diferentes vías:

a) Sobre las lenguas manejadas en las interfaces de los portales (véase 1, CCn[1]) Boelhouwer/Dykstra/Sijens (2017: 759) indican que los 37 portales que han analizado abarcan un total de 22 lenguas. De estos, 16 ofrecen más de una lengua de interfaz y, en estos casos, el inglés es siempre una de ellas. Por el contrario, 11 portales (29,27%) no usan el inglés como lengua de interfaz. La relación entre tipo de portal y lengua de interfaz arroja datos significativos: los motores de búsqueda muestran predilección por el uso de una única lengua (13 de ellos solo usan una lengua de interfaz, frente a 5 redes y 1 repositorio); portales con 2 lenguas de interfaz muestran índices semejantes (4 redes, 3 agregadores y 2 repositorios usan 2 lenguas); y entre los que tienen 3 o más lenguas de interfaz destacan los agregadores con 7 representantes. En cuanto al grado de autonomía de los recursos contemplados en los portales y plataformas (véase 1, CCl[5]) y su accesibilidad como recursos independientes, Boelhouwer/Dykstra/Sijens (2017: 760) señalan que solo 1 red y 6 agregadores, de entre todos los analizados, no cumplen este requisito. Esto supone menos de un 20% de los portales analizados.

b) Los portales virtuales así como muchas plataformas y nuevos recursos como redes semánticas (*FrameNet* y *WordNet*) parecen proceder de modo similar en lo relativo al estatus de las lenguas y el grado de profundidad de análisis. Entre los portales que tienen como objetivo plasmar

24. Se define en su página informativa como un multilingual encyclopedic dictionary. Resulta especialmente interesante por su carácter colaborativo y por integrar automáticamente información de diferentes recursos, tales como WordNet, OmegaWiki, Wiktionary, Wikipedia, Wikidata y Open Multilingual WordNet, entre otros.

en primera instancia datos lingüísticos –no, por ejemplo, los repositorios–, se observa que la mayoría de ellos son recursos *de facto* plurilingües con consultas ceñidas a pares de lenguas concretos, como se observa, por ejemplo, en el portal de *Larousse* o en *Reverso*[25]:

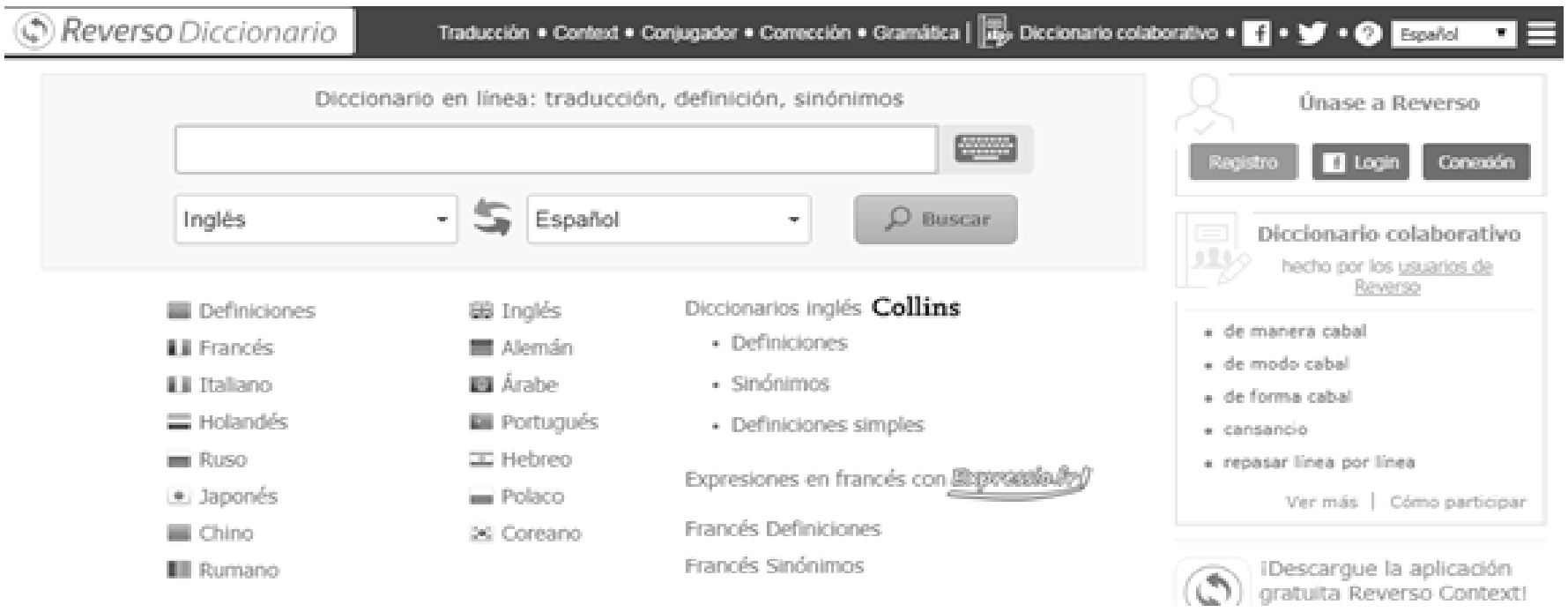

Imagen 12: *Reverso*

En nuestra opinión, resultan más excepcionales –también más actuales y a menudo de corte colaborativo (Abel /Meyer, 2013)–, aquellos que son multilingües, como por ejemplo *Omegawiki*:

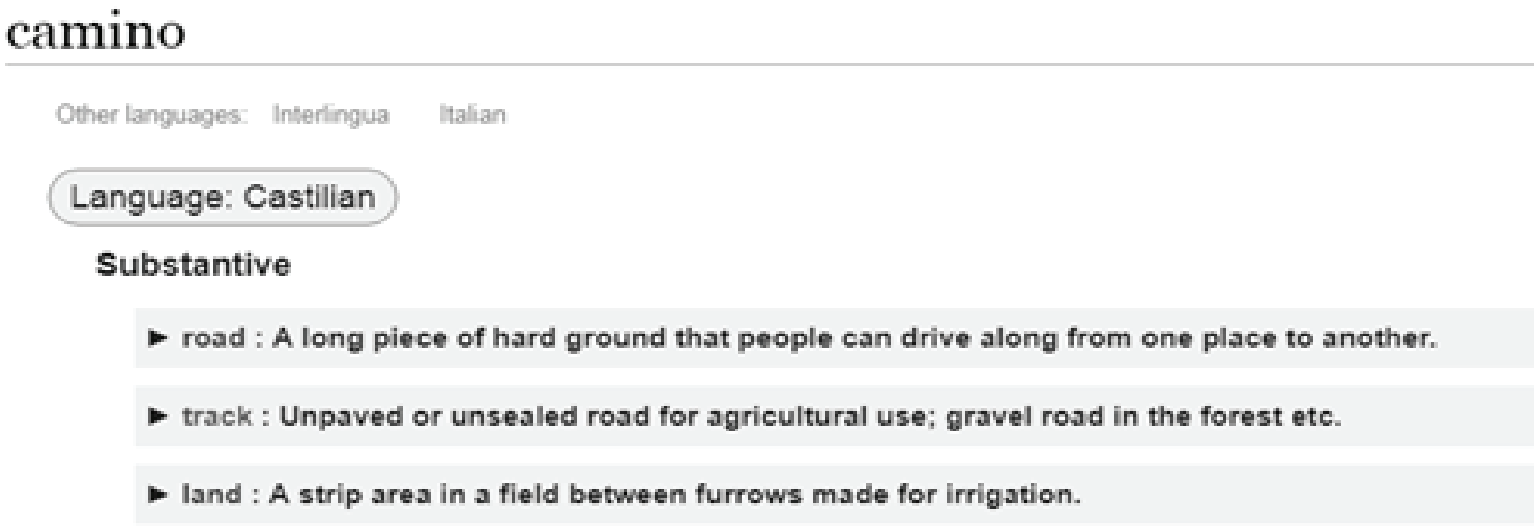

Imagen 13: Resultado de la búsqueda de 'camino' en español

25. *Reverso* aporta diccionarios generales y especializados, servicios de traducción automática, un conjugador, una gramática, etc., y permite, además, búsquedas en corpora paralelos (el así llamado diccionario contextual), en el diccionario colaborativo, en enciclopedias, entre otros.

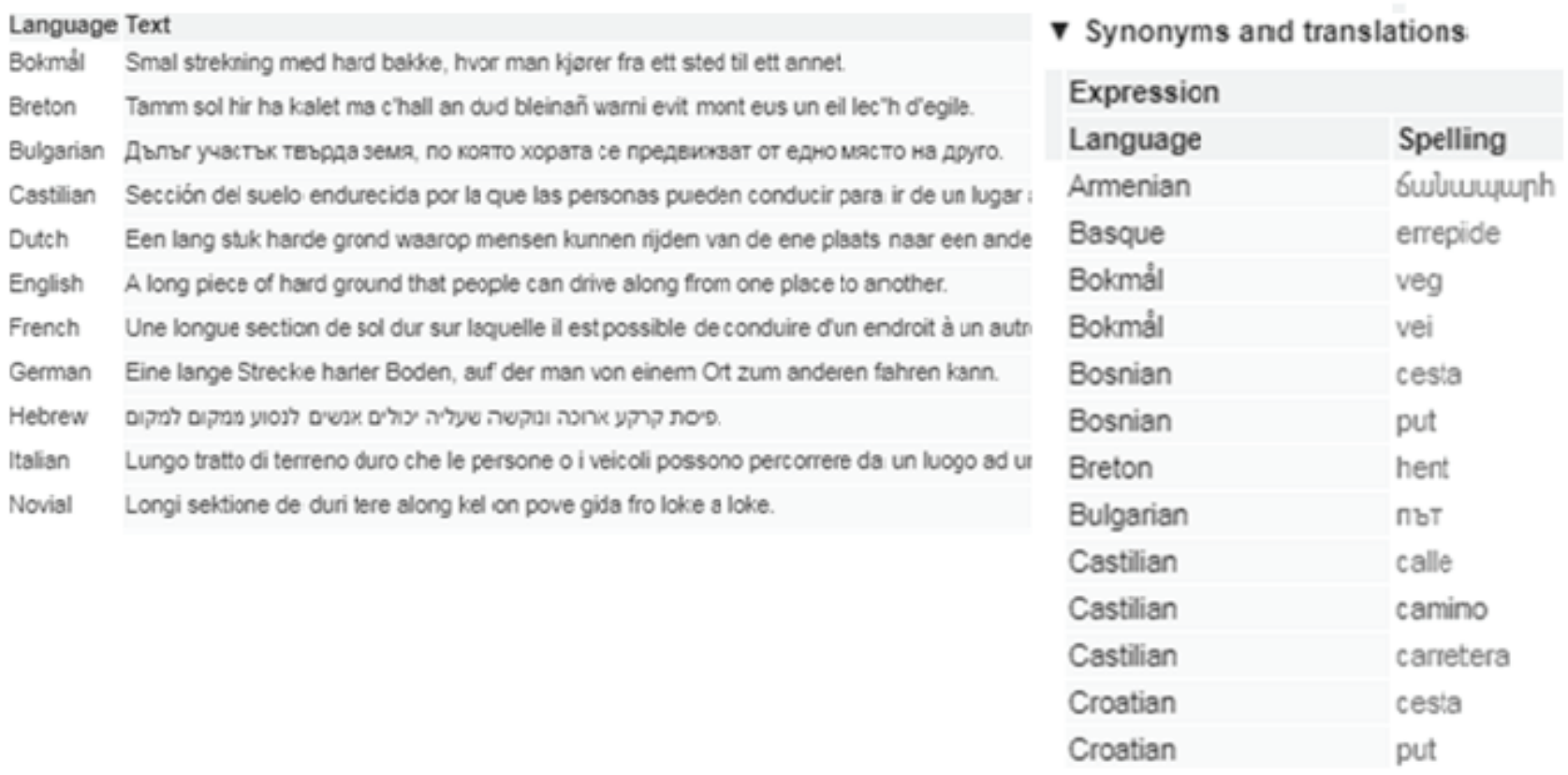

| Language | Text |
|---|---|
| Bokmål | Smal strekning med hard bakke, hvor man kjører fra ett sted til ett annet. |
| Breton | Tamm sol hir ha kalet ma c'hall an dud bleinañ warni evit mont eus un eil lec'h d'egile. |
| Bulgarian | Дълъг участък твърда земя, по която хората се предвижват от едно място на друго. |
| Castilian | Sección del suelo endurecida por la que las personas pueden conducir para ir de un lugar a |
| Dutch | Een lang stuk harde grond waarop mensen kunnen rijden van de ene plaats naar een ande |
| English | A long piece of hard ground that people can drive along from one place to another. |
| French | Une longue section de sol dur sur laquelle il est possible de conduire d'un endroit à un autr |
| German | Eine lange Strecke harter Boden, auf der man von einem Ort zum anderen fahren kann. |
| Hebrew | פיסת קרקע ארוכה ונקשה שעליה יכולים אנשים לנסוע ממקום למקום. |
| Italian | Lungo tratto di terreno duro che le persone o i veicoli possono percorrere da un luogo ad u |
| Novial | Longi sektione de duri tere along kel on pove gida fro loke a loke. |

▼ Synonyms and translations

| Expression | |
|---|---|
| Language | Spelling |
| Armenian | ճանապարհ |
| Basque | errepide |
| Bokmål | veg |
| Bokmål | vei |
| Bosnian | cesta |
| Bosnian | put |
| Breton | hent |
| Bulgarian | път |
| Castilian | calle |
| Castilian | camino |
| Castilian | carretera |
| Croatian | cesta |
| Croatian | put |

Imagen 14: Resultado de la acepción de 'camino' (road) en *Omegawiki* (acortado)

## 3.3 Índice interlingüístico y multilateralidad

Atendiendo a los diferentes tipos de portales y plataformas –y al igual que ya sucedía con los diccionarios– resulta complejo extraer conclusiones tipológicas generales en lo concerniente a este apartado (vid. 1, CCI[3]). Sí se observan, como cabía esperar, diferentes tipos. Así, al igual que el *Wiktionary*, *Dicovia.com*[26] (Imagen 15), o *Dicts.info* (Imagen 16) operan con formas (*wordforms*) como índice interlingüístico:

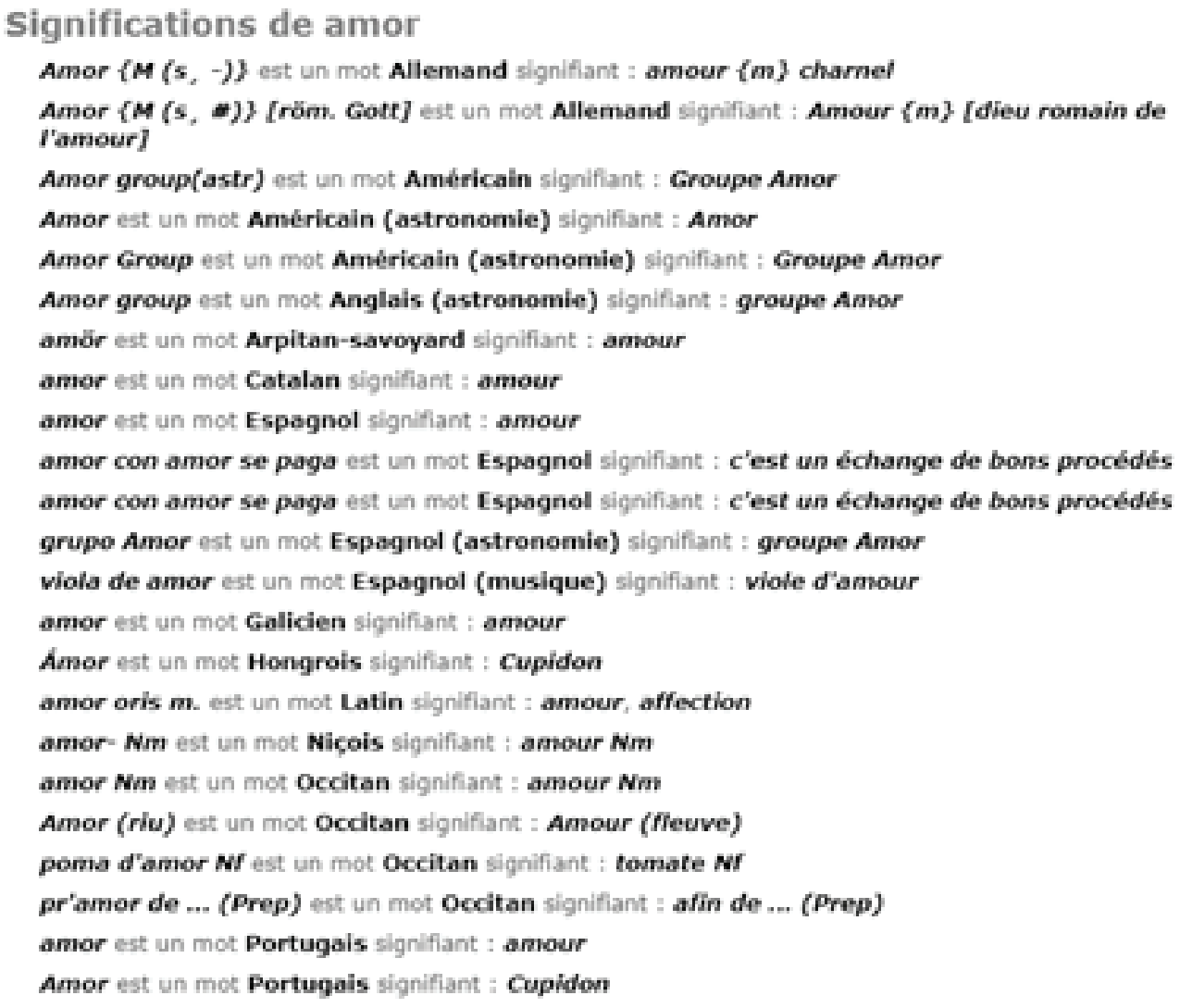

**Significations de amor**

***Amor {M (s, -)}*** est un mot **Allemand** signifiant : ***amour {m} charnel***
***Amor {M (s, #)} [röm. Gott]*** est un mot **Allemand** signifiant : ***Amour {m} [dieu romain de l'amour]***
***Amor group(astr)*** est un mot **Américain** signifiant : ***Groupe Amor***
***Amor*** est un mot **Américain (astronomie)** signifiant : ***Amor***
***Amor Group*** est un mot **Américain (astronomie)** signifiant : ***Groupe Amor***
***Amor group*** est un mot **Anglais (astronomie)** signifiant : ***groupe Amor***
***amör*** est un mot **Arpitan-savoyard** signifiant : ***amour***
***amor*** est un mot **Catalan** signifiant : ***amour***
***amor*** est un mot **Espagnol** signifiant : ***amour***
***amor con amor se paga*** est un mot **Espagnol** signifiant : ***c'est un échange de bons procédés***
***amor con amor se paga*** est un mot **Espagnol** signifiant : ***c'est un échange de bons procédés***
***grupo Amor*** est un mot **Espagnol (astronomie)** signifiant : ***groupe Amor***
***viola de amor*** est un mot **Espagnol (musique)** signifiant : ***viole d'amour***
***amor*** est un mot **Galicien** signifiant : ***amour***
***Ámor*** est un mot **Hongrois** signifiant : ***Cupidon***
***amor oris m.*** est un mot **Latin** signifiant : ***amour, affection***
***amor- Nm*** est un mot **Niçois** signifiant : ***amour Nm***
***amor Nm*** est un mot **Occitan** signifiant : ***amour Nm***
***Amor (riu)*** est un mot **Occitan** signifiant : ***Amour (fleuve)***
***poma d'amor Nf*** est un mot **Occitan** signifiant : ***tomate Nf***
***pr'amor de ... (Prep)*** est un mot **Occitan** signifiant : ***afin de ... (Prep)***
***amor*** est un mot **Portugais** signifiant : ***amour***
***Amor*** est un mot **Portugais** signifiant : ***Cupidon***

Imagen 15: *Dicovia.com*

26. *Dicovia.com* es un motor de búsqueda que permite encontrar una traducción en unos 200 diccionarios.

| | |
|---|---|
| **amor (m)** amor | **amor** amor (m) |
| **amor (m)** amor (m) | **amor (m)** amor (m) |
| **amor (m); cariño (m)** amor | **amor** amor (m); cariño (m) |
| **hacer el amor** amar; fazer amor | **amar; fazer amor** hacer el amor |
| **tener sexo; hacer el amor** transar | **fazer amor** hacer el amor |
| **hacer el amor** fazer amor | **amora (f)** mora (f) |
| **zarzamora (f)** amoreira (f) | **amoreira (f)** zarzamora (f) |

Imagen 16: *Dicts.info*[27]

A diferencia de estos, la plataforma *BabelNet* recurre a conceptos (Imagen 17), al igual que redes semánticas como *WordNet*, y permite, además, consultas multilaterales desde cualquiera de las lenguas del recurso (Imagen 18):

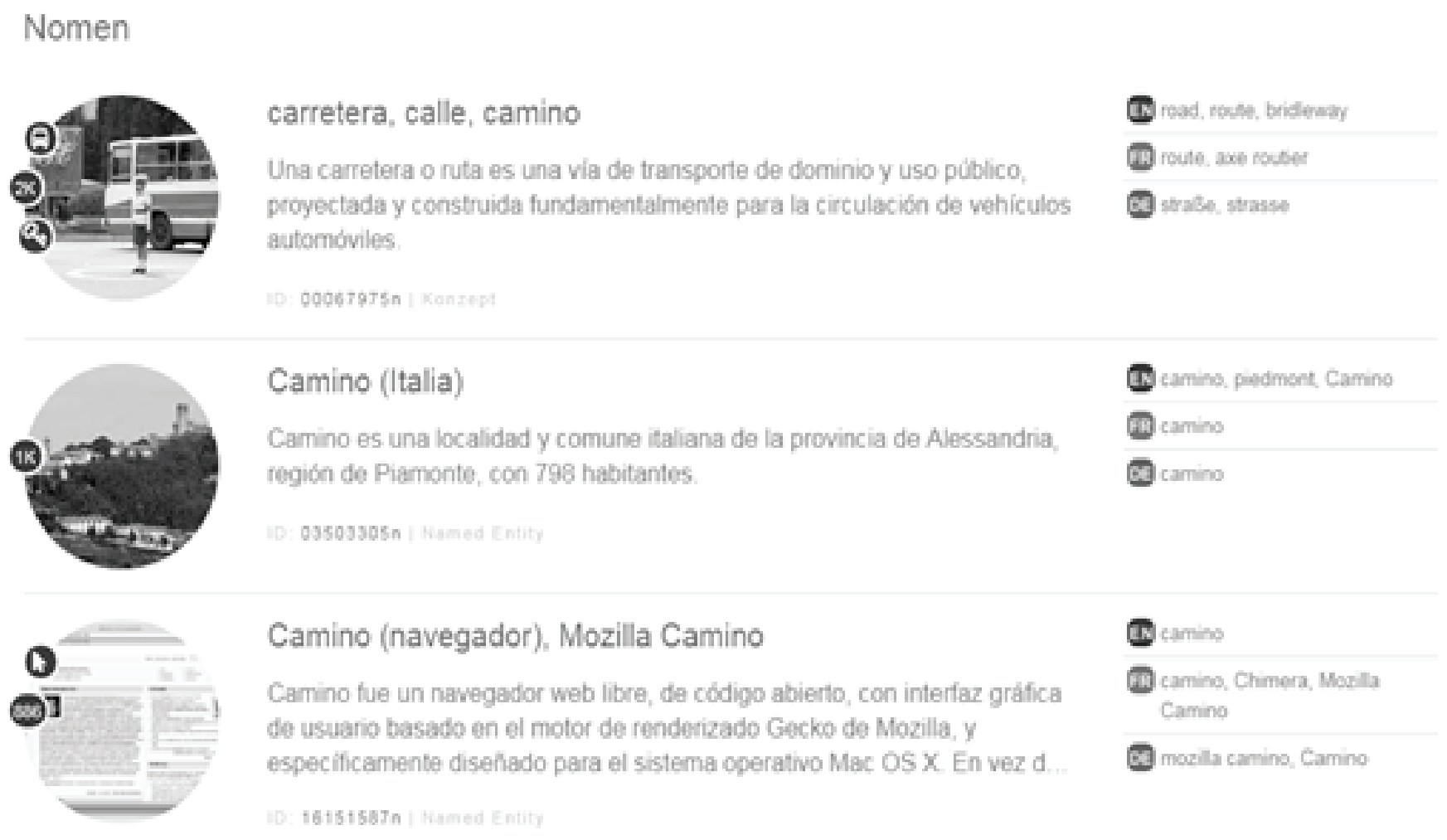

Imagen 17: Resultados de la consulta de 'camino' en *Babelnet*

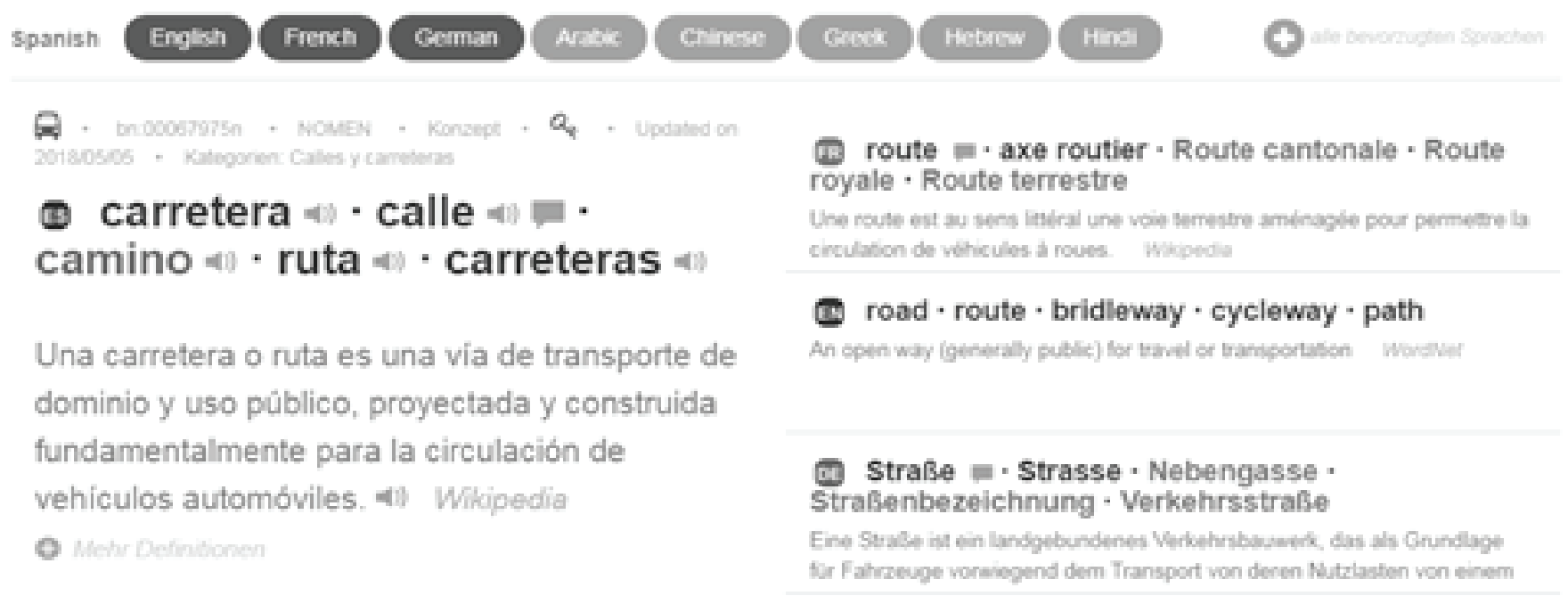

Imagen 18: Resultados de la consulta de la primera acepción de 'camino' en *Babelnet*

27. Obsérvese la inclusión de 'zarzamora' en los resultados, si bien la consulta realizada fue 'amor'.

### 3.4 A modo de resumen

A lo largo de los párrafos anteriores, se ha presentado una serie de portales que indexan recursos que aportan información sobre diferentes lenguas. Muchos de ellos son recursos plurilingües, que indexan diccionarios bilingües. Permiten consultas monolingües y bilingües en los diferentes diccionarios indexados, pero en menos ocasiones ofrecen información multilingüe, alineada y comparable con una misma estructura de acceso entre más de dos lenguas. Falta aún un análisis exhaustivo de los portales en relación con el papel que las lenguas descritas desempeñan en los mismos, más allá del estudio de Boelhouwer/Dykstra/Sijens (2017; véase 3.2.), que presenta datos sobre las lenguas manejadas en las interfaces de los portales.

## 4. Para finalizar

Para caracterizar un recurso como *multilingüe* frente a uno *plurilingüe* no podemos contemplar únicamente el criterio del número de lenguas, puesto que este –siendo superior a dos– no es distintivo. Los rasgos determinantes son, pues, a) la cantidad y la calidad de la información en términos interlingüísticos que se aportan en un tipo y otro, b) las opciones múltiples de acceso y búsqueda de información –en parámetros de unilateralidad, bilateralidad, multilateralidad y *cross-lingual*–, c) la polifuncionalidad del recurso y d) la interconexión de los datos aportados. Entendemos, por tanto, que un recurso multilingüe adjudica a todas las lenguas el mismo estatus jerárquico, presenta una relación interlingüística todas><todas, una interconexión de datos modular alineada (en su defecto interconexión sólo lema) y una interconexión relacional de los datos. Frente a éste, el término de *plurilingüe* describe en este trabajo recursos que presentan más de dos lenguas, si bien el peso concedido a las lenguas no es equilibrado.

Asimismo, teniendo en cuenta la diversidad de recursos lexicográficos electrónicos existentes, se precisa una propuesta tipológica global en el campo de la lexicografía multilingüe, ya no solo en cuanto a la delimitación de los recursos según su carácter multilingüe o plurilingüe, tal y como aquí se postula, sino también atendiendo a las notables diferencias existentes entre diccionarios y portales –y entre estos últimos entre sí; pero, además, hoy en día contamos con nuevos recursos de aplicación lexicográfica, tales como las redes semánticas, las plataformas de recursos integrados, etc. que retan a encontrarles cabida, o no, en una clasificación de los recursos electrónicos multilingües.

# Referencias bibliográficas

## Literatura

## Diccionarios, portales y plataformas